\documentstyle[11pt,conf_iap,epsf,psfig]{article}
\newcommand{\etal}{{\it et al.}}
\newcommand{\mnras}{{\em Mon. Not. R. astr. Soc}}
\newcommand{\an}{   {\em Astron. Nach.}}
\newcommand{\aeta}{ {\em Astron. Astrophys.}}
\newcommand{\aetas}{{\em Astron. Astrophys. Suppl. Ser.}}
\newcommand{\physlettb}{{\em Phys. Lett. B}}
\newcommand{\physrev}{{\em Phys. Rev.}}
\renewcommand{\etal}{{\it et al.}}

\newcommand{\araa}{ {\em Ann. Rev. Astron. Astrophys.}}
\def\mystrut{ \protect \rule[-0.3cm]{0cm}{0.8cm} }

\begin{document}

\heading{THE TYCHO DATABASE \\
 AS A CONTROL MICROLENSING EXPERIMENT}

\photo{ }

\author{T. DIMEO$^{1}$, D. VALLS--GABAUD$^{1,2}$, E. J. KERINS$^{1}$ }
       {$^1$ Observatoire de Strasbourg, 11 Rue de l'Universit\'e, 67000 Strasbourg, France\\
        $^2$ Royal Greenwich Observatory, Madingley Road, Cambridge CB3 0EZ,
 United Kingdom\\
Email : {\tt dimeo/dvg/kerins@astro.u-strasbg.fr} }

\bigskip

\begin{abstract}{\baselineskip 0.4cm 
The amount of dark matter in the solar
neighbourhood is a poorly constrained quantity which is nevertheless
used to normalise the properties of the dark halo. While the current surveys
towards the LMC, SMC and the Galactic bulge probe the dark matter
content at large distances from the Sun, the HIPPARCOS satellite provides
a unique way to constrain the dark matter within about 1 kpc. The TYCHO program
observed about a million stars in two passbands down to a magnitude
of about 12, all over the sky. The large number of observations taken during 
the 3 year mission make  such a catalogue a useful probe to validate the
microlensing technique to detect baryonic compact objects, since the results
can be compared to the local dynamical constraints.  
  We predict here that about 0.1  events  could 
be present in the TYCHO database, but after extensive 
 Monte Carlo simulations that
take into account both  temporal sampling and photometric errors, we
find that the
efficiency is only 9\% for typical events of 15 days, reducing the expected
number of events to 0.01. 
}
\end{abstract}

\section{Dark matter in the solar neighbourhood}

The detection, by means of the gravitational amplification of light, of
invisible objects has ceased to be a dream \cite{Liebes64} 
to become a reality  thanks to the
efforts of a number of teams (see \cite{BP96araa} for a review). This
technique may shed new light on the nature of dark matter, and is 
totally independent of more or less direct methods of detection. Hence
it seems important to ``calibrate'' it by performing ``control''
experiments in volumes where the dynamical amount of dark matter is
thought to be reasonably well constrained by a variety of methods, between
about 46 to 90 $M_\odot$ pc$^{-2}$ (to be compared to the observed density in gas and
stars which amounts to about 40 $M_\odot$ pc$^{-2}$). This
is also important because so far the experiments towards both the LMC 
and the Galactic bulge have given surprising results. Here
we suggest using the TYCHO database to test for the presence of microlensing
events in the solar neighbourhood. This is the best control experiment
to calibrate not only the other experiments, but also the theoretical
predictions which use, as a normalisation, this local amount of dark matter
 \cite{Flores88}.

\begin{center}
{\bf Table 1.} Basic 3-component model for the solar neighbourhood
\end{center}
\begin{center}
\begin{tabular}{|l| r@{}l cc|ccc|cc|}
\hline
Component \mystrut  & 
\multicolumn{2}{c}{$\rho_o$ }& H & h & $U$ & $V$ & $W $ & 
V$_{\rm rot}$ & $a$ \\
  &
\multicolumn{2}{c}{  [M$_\odot$ pc$^{-3}$]}&
      [pc] &  [kpc] & [km/s] & [km/s] & [km/s] & [km/s] &  [kpc] \\
\hline 
Thin disc \mystrut & \hspace*{5mm} 0&.05  & 300 & 3.0 & 35 & 25 & 25 & 220 & --  \\
Thick disc \mystrut & \hspace*{5mm} 0&.002 & 800 & 3.0 & 67 & 51 & 40 & 167 & -- \\
Halo  \mystrut     & \hspace*{5mm} 0&.01  &  -- & --  &  156 & 156 & 156 & 0 & 5.0 \\
\hline
\end{tabular}
\end{center}

\section{Predicting microlensing events in the TYCHO database}

The TYCHO experiment has used the star mapper of the HIPPARCOS satellite
to produce a photometric and astrometric catalogue over the full sky 
of about a million stars down to a magnitude of about 12 \cite{Hog92}.
 This is the first catalogue to contain measures of both
northern and southern stars with the same instrument and a well defined
colour system (quasi $B$ and $V$), with over a hundred measures per star spanning the 36 months
of data acquision by HIPPARCOS \cite{Hog92} \cite{Scales92}
\cite{Grossman95}. The TYCHO experiment is thus an ideal database to look for
possible microlensing signatures of lenses within the solar neighbourhood.

We have used an improved Monte Carlo code \cite{Kerins95} to estimate the microlensing properties
of the solar neighbourhood. The 
 thick and thin  discs were modelled as double exponential
discs, with properties as given in Table 1, derived from recent surveys
\cite{Ojha96} \cite{Denis96}. The lenses are randomly
distributed following the 3 components given in Table 1, while the
sources are assumed to be essentially in the thin disc. Given the shallow
depth of TYCHO, there was no need to include the bulge and spheroid components.

The are two main sources of uncertainty in the present analysis, in
addition to the uncertainties in the dynamical properties of the sources.

\indent \hspace*{2mm} $\Diamond$ First, the actual optical depth  depends on the luminosity function of the
sources. The preliminary HR Diagram from HIPPARCOS
\cite{Mac95} markedly  shows a maximum at about +4, along with a densely
populated giant branch and clump at about +1. Given that the HIPPARCOS
data are not yet available, we used the INCA catalogue \cite{Egret92} 
restricting the analysis to the 1612 stars whose parallaxes are quoted with
less than 20\% error. The depth is comprised between 0.4 and 1 kpc with
 this approximate luminosity
function at a limiting magnitude of 12.

\indent \hspace*{2mm} $\Diamond$ Second, the patchiness of the extinction is a severe problem that limits the
accuracy of the predictions. We assume a uniform extinction of 
$A_V$ = 1 mag kpc$^{-1}$ within a disc of thickness
 $H_z$=100 pc, which is a bit larger than the
usual 0.8 mag kpc$^{-1}$, but in better agreement with the extinction maps of
the solar neighbourhood \cite{Neckel80}. The effect of extinction can be
noted on Figs. 1 and 2, where the maximum optical depth and rate are not on
the plane, but slightly above and below. Note also that the strong dependence with
longitude is due to the fact that the sources are in the thin disc, and so
the density is  smaller in the anticentre direction.

The microlensing rate, averaged over the sky, is 3.6 10$^{-8}$ events/year for 0.5 $M_\odot$ lenses. Since during the 3 years of observations an average of
 100 observations were taken per star, for a target number of a million
stars, a total of 0.11 events can be expected\footnote{This is in sharp
contrast to a previous analysis \cite{Hog95} which only considered 
lenses arising from the halo.}. Microlensing time scales averaged over the Earth's motion are given in
Fig. 3, where both the Sun's motion with respect to the LSR and the orbital
motion of the Earth were taken into account. They span the range from 8 to 50
days for a 0.5 $M_\odot$ lens.

Monte Carlo error estimates for the optical depth, rate and time scales are 3--4 \%,
5--6\% and 6--7\% respectively, hence the difference between neighbouring
contours should not be taken too literally. Also the small spikes that
appear at high $|b|$ are a signature of the Monte Carlo noise only.

\begin{figure}[p]
\epsfxsize=15.0cm
\epsfbox{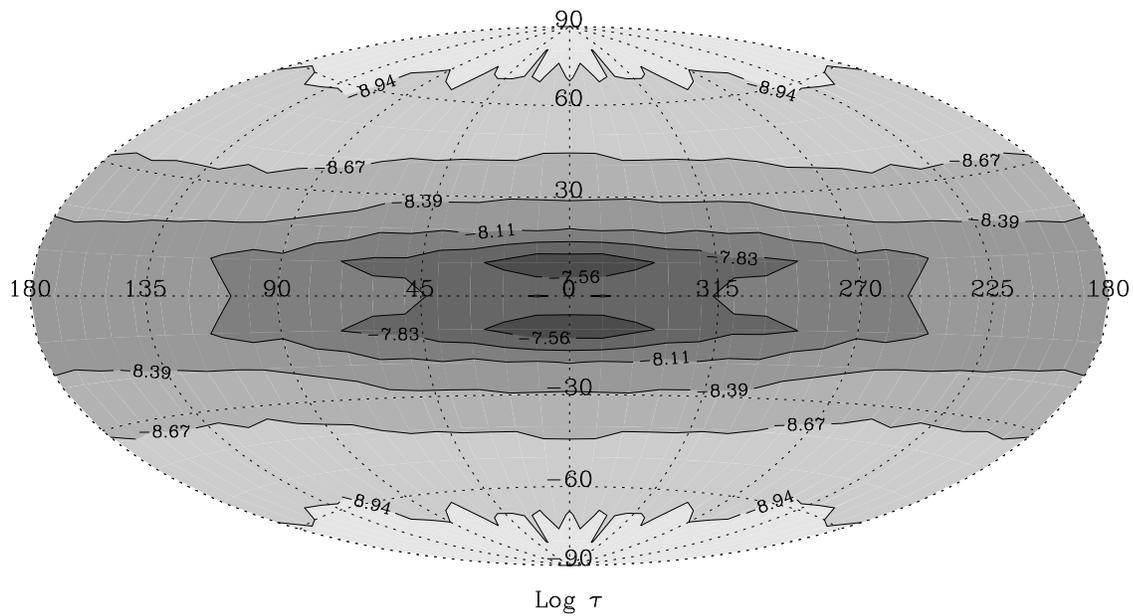}
\caption[]{Predicted angular dependence of the microlensing optical
depth for the TYCHO database.}
\label{figtau}
\end{figure}
\samepage
\begin{figure}[p]
\epsfxsize=15.0cm
\epsfbox{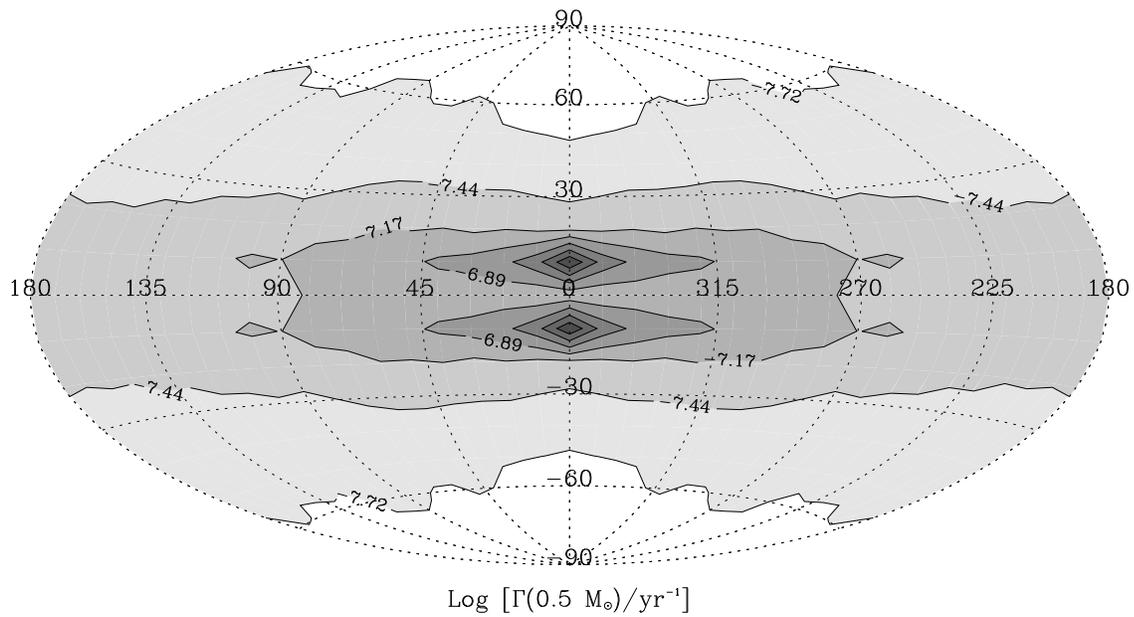}
\caption[]{Predicted TYCHO microlensing event rate across the sky. Note the
effects of extinction across the galactic disc.}
\label{figrate}
\end{figure}

\clearpage

\begin{figure}[t]
\epsfxsize=15.0cm
\epsfbox{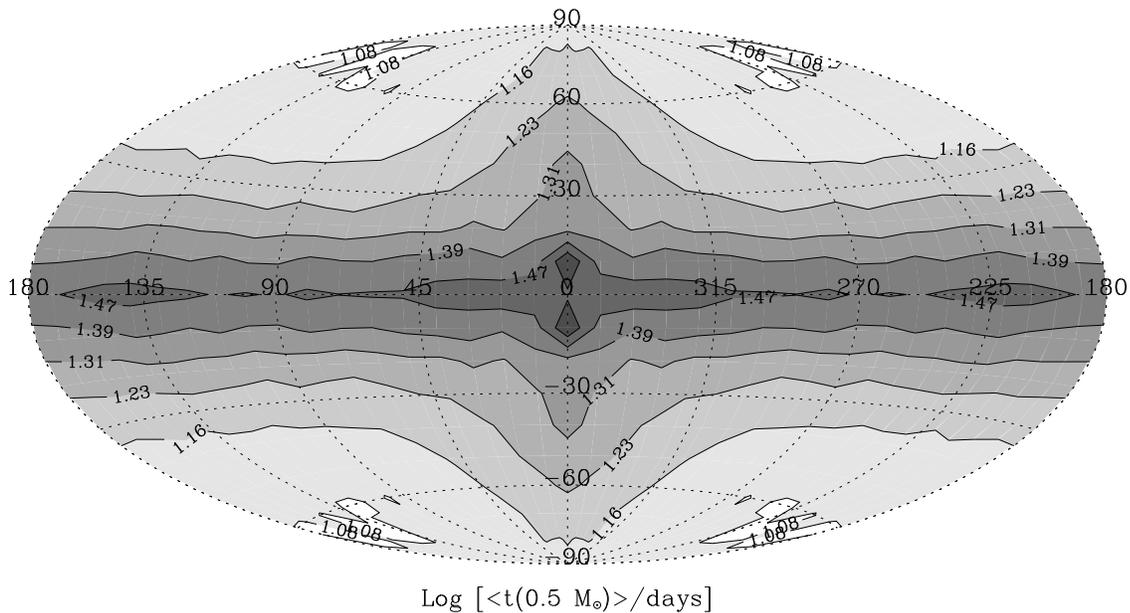}
\caption[]{Predicted microlensing time scales for TYCHO, ranging
from 8 days  to about 40 days, for a 0.5 $M_\odot$ lens.}
\label{figtime}
\end{figure}
\samepage
\begin{figure}[b]
\centerline{
\psfig{file=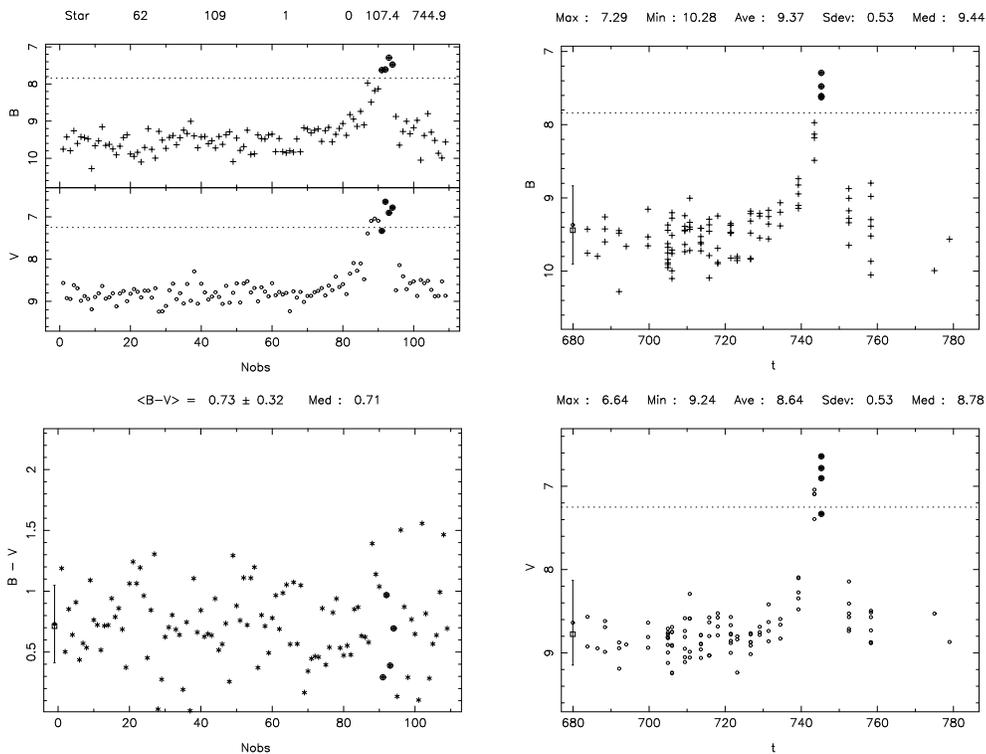,width=13.0cm,angle=-90.}}
\caption[]{Simulated microlensing event with the TYCHO time sampling and 
photometric errors. The upper left panel shows the photometry vs. 
observation number; the upper right panel the $B$ light curve, the
lower left panel the $B-V$ simulated colours, and the lower right
panel gives the $V$ lightcurve. The dotted line indicates the 3$\sigma$
level above the median magnitude in each passband.}
\label{figsim}
\end{figure}

\clearpage


However, one has to take into account the effects of sampling and the
photometric errors. We used simple approximations based on the preliminary
analysis of TYCHO data \cite{Scales92} \cite{Grossman95} to make extensive Monte Carlo simulations
to derive the efficiency of TYCHO as a microlensing experiment (see Fig. 4). This gave an efficiency
of about 9\% for events of 10 days, and up to 15\% for durations of a couple
of months. Hence the efficiency decreases 
the  expected number of events down to about 0.01.
Even though this is disappointing, we argue that the experiment
{\sl should} be carried out, given the unexpected results obtained by the
current microlensing surveys of the LMC and Galactic bulge.

\section{Conclusions}
We have shown that the TYCHO database could be used as a control
microlensing experiment to probe the local dark matter content in the
solar neighbourhood. The same methodology as the current microlensing 
surveys can be used to set limits on the local dark matter probed by
the microlensing experiments, which can then be compared to the limits
obtained by dynamical methods. The low rate expected for the nearby 
stars probed by TYCHO is somehow
compensated by the large number of stars in the survey, and by the large
number of observations per star. The sampling efficiency is best
suited to long events (months), although very short time scales
(20 to 120 minutes) are also probed. Even though the predicted number 
of events detectable in the final database (when available in 1997) is
very small, we argue that it is important to carry out this experiment to validate the basic microlensing
methodology.

In the future, the full analysis of the entire  TYCHO batabase
 and the astrometric satellite projects such
 as DIVA \cite{DIVA}, or GAIA and ROEMER \cite{Hog95}, 
monitoring some 10$^8$ stars down to V=16.5 (and hence
probing some 6 kpc), could  be used to validate the
microlensing technique to probe the dark matter content of the 
solar neighbourhood.

\acknowledgements{We are grateful to J.L. Halbwachs for useful
discussions on the TYCHO database.}

\vfill

\end{document}